\documentclass[11pt,twoside]{article}
\usepackage{./asp2014}

\aspSuppressVolSlug
\resetcounters

\begin{document}

\title{The Impact of Bulges on the Radial Distribution of Supernovae in Disc Galaxies}
\author{A.~A.~Hakobyan,$^1$ A.~G.~Karapetyan,$^1$ L.~V.~Barkhudaryan,$^1$ G.~A.~Mamon,$^2$ D.~Kunth,$^2$
        A.~R.~Petrosian,$^1$ V.~Adibekyan,$^3$ L.~S.~Aramyan,$^1$ and M.~Turatto$^4$
\affil{$^1$Byurakan Astrophysical Observatory, Byurakan, Armenia;
       \email{hakobyan@bao.sci.am, karapetyan@bao.sci.am}\\
       $^2$Institut d'Astrophysique de Paris, Paris, France;\\
       $^3$Instituto de Astrof\'{i}sica e Ci\^{e}ncia do Espa\c{c}o, Porto, Portugal;\\
       $^4$Osservatorio Astronomico di Padova, Padova, Italy}}

\paperauthor{A.~A.~Hakobyan}{hakobyan@bao.sci.am}{}{Byurakan Astrophysical Observatory}{}{Byurakan}{}{}{Armenia}
\paperauthor{A.~G.~Karapetyan}{karapetyan@bao.sci.am}{}{Byurakan Astrophysical Observatory}{}{Byurakan}{}{}{Armenia}
\paperauthor{L.~V.~Barkhudaryan}{}{}{Byurakan Astrophysical Observatory}{}{Byurakan}{}{}{Armenia}
\paperauthor{G.~A.~Mamon}{}{}{Institut d'Astrophysique de Paris}{}{Paris}{}{}{France}
\paperauthor{D.~Kunth}{}{}{Institut d'Astrophysique de Paris}{}{Paris}{}{}{France}
\paperauthor{A.~R.~Petrosian}{}{}{Byurakan Astrophysical Observatory}{}{Byurakan}{}{}{Armenia}
\paperauthor{V.~Adibekyan}{}{}{Instituto de Astrof\'{i}sica e Ci\^{e}ncia do Espa\c{c}o}{}{Porto}{}{}{Portugal}
\paperauthor{L.~S.~Aramyan}{}{}{Byurakan Astrophysical Observatory}{}{Byurakan}{}{}{Armenia}
\paperauthor{M.~Turatto}{}{}{Osservatorio Astronomico di Padova}{}{Padova}{}{}{Italy}

\begin{abstract}
  We present the results of the analysis of the impact of bulges on the radial distributions
  of the different types of supernovae (SNe) in the stellar discs of
  host galaxies with various morphologies.
  We find that in Sa--Sm galaxies, all core-collapse (CC)
  and vast majority of SNe Ia belong to the
  disc, rather than the bulge component.
  The radial distribution of SNe Ia in S0--S0/a galaxies is
  inconsistent with their distribution in Sa--Sm hosts,
  which is probably due to the contribution
  of the outer bulge SNe Ia in S0--S0/a galaxies.
  The radial distributions of both types of SNe are similar
  in all the subsamples of Sa--Sbc and Sc--Sm galaxies.
  These results confirm that the old bulges of Sa--Sm galaxies are
  not significant producers of Type Ia SNe, while the bulge populations
  are significant for SNe Ia only in S0--S0/a galaxies.
\end{abstract}

\section*{Summary}

In \citet{2016MNRAS.456.2848H}, using a well-defined and homogeneous
sample of SNe and their host galaxies from the coverage of
Sloan Digital Sky Survey \citep[]{2012A&A...544A..81H,2014MNRAS.444.2428H}, we analysed the impact
of bulges on the radial distributions of the different types of SNe
in host galaxies with various morphologies.
Our sample consists of 419 nearby (${\leq {\rm 100~Mpc}}$),
low-inclination ($i \leq 60^\circ$), and morphologically non-disturbed S0--Sm galaxies,
hosting 500 SNe in total.

All the results that we summarize below concerning
the impact of bulges on the radial distributions of SNe in hosts
can be explained considering that the old bulges of Sa--Sm galaxies are
not significant producers of Type Ia SNe, while the bulge populations
are significant for SNe Ia only in S0--S0/a galaxies.

\begin{itemize}
\item In Sa--Sm galaxies, all CC and the vast majority of Type Ia SNe belong to the disc,
      rather than the bulge component.
      This result suggests that the rate of SNe Ia in spiral galaxies is dominated by
      a relatively young/intermediate progenitor population \citep[e.g.][]{2005A&A...433..807M,2011Ap.....54..301H,2011MNRAS.412.1473L}.
      In addition, this observational fact makes the deprojection of galactocentric radii
      of both types of SNe a key point in the statistical
      studies of their distributions \citep[][]{2008Ap.....51...69H,
      2009A&A...508.1259H,2013Ap&SS.347..365N}.
\item The radial distribution of Type Ia SNe in S0--S0/a galaxies is
      inconsistent with that in Sa--Sm hosts.
      This inconsistency is mostly attributed to the contribution by SNe Ia
      in the outer bulges of S0--S0/a galaxies. In these hosts, the relative
      fraction of bulge to disc SNe Ia is probably changed in comparison
      with that in Sa--Sm hosts, because the progenitor population from
      the discs of S0--S0/a galaxies should be much lower due to the lower
      number of young/intermediate stellar populations.
\item The inner fractions ($R_{\rm SN} \leq 0.3\,R_{25}$,
      where the bulge stars are tipically located) of Type Ia and CC SNe
      are not statistically different one
      from another in morphologically non-disturbed spiral
      hosts.\footnote{{\footnotesize The $R_{25}$ is the SDSS $g$-band
      $25^{\rm th}$ magnitude isophotal semimajor axis of SN host galaxy.}}
      This result additionally confirms that the old bulges of Sa--Sm galaxies are
      not significant producers of Type Ia SNe.
\end{itemize}

{\footnotesize \acknowledgements This work was supported by the RA MES State Committee of Science,
in the frames of the research project number 15T--1C129.
This work was made possible in part by a research grant from the
Armenian National Science and Education Fund (ANSEF) based in New York, USA.
V.A. acknowledges the support from FCT through Investigador FCT contracts of reference IF/00650/2015.
V.A. also acknowledges the support from Funda\c{c}\~ao para a Ci\^encia e Tecnologia (FCT) through national funds and from FEDER through COMPETE2020 by the following grants UID/FIS/04434/2013 \& POCI-01-0145-FEDER-007672, PTDC/FIS-AST/7073/2014 \& POCI-01-0145-FEDER-016880 and PTDC/FIS-AST/1526/2014 \& POCI-01-0145-FEDER-016886.}



\end{document}